\documentclass{vldb}
\usepackage{graphicx}
\usepackage{balance}  
\usepackage[hidelinks]{hyperref}
\usepackage{float}
\usepackage{pslatex}
\usepackage{xcolor}
\usepackage{enumitem}
\usepackage[numbers]{natbib}
\usepackage[skip=0.9\baselineskip,font=bf]{caption}
\usepackage[utf8]{inputenc}

\vldbTitle{SciLens News Platform: A System for Real-Time Evaluation of News Articles}
\vldbAuthors{Angelika Romanou, Panayiotis Smeros, Carlos Castillo, Karl Aberer}
\vldbDOI{https://doi.org/10.14778/3415478.3415521}
\vldbVolume{13}
\vldbNumber{12}
\vldbYear{2020}

\newcommand{\rurl}[1]{\href{http://#1}{#1}}

\begin{document}
\sloppy

\pagestyle{empty}


\title{SciLens News Platform: A System for Real-Time Evaluation of News Articles}



%
%
%
%


\author{
Angelika Romanou$^\dagger$\hspace{0.10in}
Panayiotis Smeros$^\dagger$\hspace{0.10in}
Carlos Castillo$^\ddagger$\hspace{0.10in}
Karl Aberer$^\dagger$
\medskip
\and
\affaddr{$^\dagger$École Polytechnique Fédérale de Lausanne (EPFL)}\\
\affaddr{Lausanne, Switzerland}\\
\affaddr{firstname.lastname@epfl.ch}
\and
\affaddr{$^\ddagger$Universitat Pompeu Fabra (UPF)}\\
\affaddr{Barcelona, Spain}\\
\affaddr{carlos.castillo@upf.edu}
}

\maketitle

\begin{abstract}
We demonstrate the SciLens News Platform, a novel system for evaluating the quality of news articles. 
The SciLens News Platform automatically collects contextual information about news articles in real-time and provides quality indicators about their validity and trustworthiness.
These quality indicators derive from i) social media discussions regarding news articles, showcasing the reach and stance towards these articles, and ii) their content and their referenced sources, showcasing the journalistic foundations of these articles.
Furthermore, the platform enables domain-experts to review articles and rate the quality of news sources.
This augmented view of news articles, which combines automatically extracted indicators and domain-expert reviews, has provably helped the platform users to have a better consensus about the quality of the underlying articles.
The platform is built in a distributed and robust fashion and runs operationally handling daily thousands of news articles.
We evaluate the SciLens News Platform on the emerging topic of \textit{COVID-19} where we highlight the discrepancies between low and high-quality news outlets based on three axes, namely their newsroom activity, evidence seeking and social engagement.
A live demonstration of the platform can be found here: \textit{\url{http://scilens.epfl.ch}}.
\end{abstract}

\section{Introduction}

\begin{figure}
  \centering
  \includegraphics[width=\linewidth]{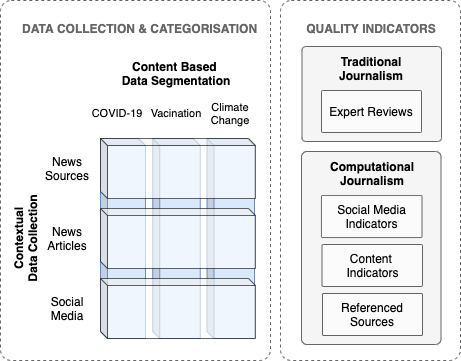}
  \caption{Overview of the SciLens News Platform, a system that collects contextual information and provides a wide range of heterogeneous quality indicators regarding news articles.}
  \label{fig:scilens}
\end{figure}

In the age of information inflation, the news is not always produced and consumed in a centralized fashion. 
Although media companies with specialized journalists are typically responsible for discovering and communicating news to the people, there are intricate ways in which this information is diffused towards the public, mostly via social media \cite{pewresearch2018}. 
The news landscape has changed radically, mainly because of: i) the instantaneous rate at which individuals publish news-worthy content, ii) the vast reachability of this content by broad audiences, and iii) the lack of regulation and quality control.
The modern news landscape consists of mainstream news outlets which are supported, complemented and often criticized by independent or alternative media channels.

However, this plethora of news sources has inevitably led to a burst of misinformation \cite{DBLP:conf/www/FernandezA18}.
Fake or fabricated stories are continuously spread among large groups of people, while credible information is often indistinguishable from misleading content.
This misleading content is a result of either detrimental and malicious behaviours (e.g., in politics) or poor interpretation and lack of domain knowledge (e.g., in science).
In the meanwhile, news companies that want to keep their users engaged, provide personalized news palettes, reinforcing preexisting media biases and empowering the influence of ``echo chambers''.

There is a wide-ranging amount of endeavours against misinformation.
These endeavours span the area of traditional journalism, with independent fact-checking initiatives (e.g., \rurl{Snopes.com}), as well as the upcoming area of data journalism \cite{DBLP:conf/www/CazalensLLMT18}, with computational methodologies against misinformation.
However, a modern news platform should be able to combine these two methodologies by i) having a robust way of processing the vast amount of information that is continuously produced, and ii) combine automatically-extracted and domain-agnostic quality indicators with the reliable but small-scale expert reviews. 

In this paper, we demonstrate the SciLens News Platform, a novel system for evaluating the quality of news articles that bridges the gap between traditional and computational journalism.
Our platform automatically extracts, stores, and displays heterogeneous quality indicators for scientific news, that were first introduced by \citet{DBLP:conf/www/Smeros0A19}. These indicators derive from i) social media discussions regarding news articles, showcasing the reach and stance towards these articles, and ii) their content and their referenced sources, showcasing the journalistic foundations of these articles.
The quality indicators are combined with expert reviews in a unified environment (Figure~\ref{fig:scilens}).
The SciLens News Platform is build in a distributed and robust fashion and runs operationally handling daily thousands of news articles.
In the following sections, we present an overview of the system and a tangible use-case on the emerging topic of \textit{COVID-19}.

\section{Related Work}
The area of computational journalism is a broad research area where results are scattered through multiple disciplines.
An extended survey on the related work of this paper is presented by \citet{DBLP:conf/www/Smeros0A19}.

As mentioned above, the traditional approach for evaluating news article quality relies on the manual work of domain experts.
Independent, non-partisan fact-checking portals perform manual content verification either on general topics (e.g., \rurl{Snopes.com}) or specific topics such as politics (e.g., \rurl{PolitiFact.com}) and science (e.g., \rurl{ScienceFeedback.co}).

Recent work demonstrates methods to automate the extraction of quality indicators.
Indicatively, \citet{DBLP:conf/www/ZhangRMASGAVLRB18} compile a detailed list of such quality indicators, while \citet{DBLP:journals/sigkdd/ShuSWTL17} introduce an approach for detecting fake news on social media based on these indicators.
\citet{doi:10.1371/journal.pone.0128193} and \citet{DBLP:conf/kdd/HassanALT17} use well-known knowledge bases like Wikipedia as ground truth for testing the validity of dubious claims, while \citet{DBLP:conf/www/PopatMSW17} describe a system that explains the news articles' stance towards such claims.

\section{System Overview}
\label{par:overview}
The SciLens Data Platform incorporates automated quality indicators for news articles (\S\ref{par:indicators}), as well as a systematic way of acquiring expert reviews (\S\ref{par:evaluation}).
The overall architecture of the system is presented in \S\ref{par:architecture}.


\subsection{Automated Quality Indicators}
\label{par:indicators}
We compute three heterogeneous sets of quality indicators, namely, content, news context, and social media indicators.
Regarding the content of a news article, we consider various well-established metrics for the quality of news such as the click-baitness of its title, the subjectivity, and readability of its body and whether it is by-lined by its author.

As for the news context of an article, we investigate the strength of the connection between this article and its primary sources of information. Thus, we consider three types of references: i) internal references within the same news outlet; many news outlets, in order to increase their user engagement, introduce such references either in ``see also'' sections or in the main body of their articles, ii) external references to potential primary sources of information (e.g., references from nation-wide news outlets to local news outlets), and iii) particularly for the case of scientific news, scientific references, i.e., references to a predefined list of academic repositories, grey-literature and peer-reviewed journals and institutional websites; as we see in our use-case (\S\ref{par:evaluation}), articles from high-quality scientific outlets are expected to have more references pointing to academic sources than articles from low-quality scientific outlets.

Finally, regarding the social media context, we measure two aspects, specifically the reach and stance towards a news article. 
Reach is measured through the proxy of social media popularity, which quantifies the impact of an article in a social media platform.
Stance, on the other hand, is the positioning of social media platform users towards an article. Stance can be positive (i.e, users support or comment on an article without expressing doubts), or negative (i.e., users question the quality of an article, or directly contradict what the article is saying). 

According to a thorough experimental evaluation which is presented by \citet{DBLP:conf/www/Smeros0A19}, the aforementioned indicators help non-expert users evaluate more accurately the quality of news articles, compared to non-experts that do not have access to these indicators.

\subsection{Expert Reviews}
\label{par:evaluation}
Along with the set of automated quality indicators, the system allows experts to annotate any article based on seven criteria: 1) Factual accuracy, 2) Scientific understanding, 3) Logic/Reasoning, 4) Precision/Clarity, 5) Sources quality, 6) Fairness, and 7) Click-baitness on a \textit{Likert Scale}\footnote{\url{https://en.wikipedia.org/wiki/Likert_scale}}, from very low quality to very high quality. 
These are common criteria used in state-of-the-art fact-checking portals like \rurl{ScienceFeedback.co}.

Based on these evaluation scores, the system computes a weighted, time-sensitive average and displays a final score of the criteria for each article. 
Optionally, expert users can provide with extensive free-text reviews about the news articles, which are also displayed to the non-expert users.


\begin{figure}[t]
  \centering
  \includegraphics[width=\linewidth]{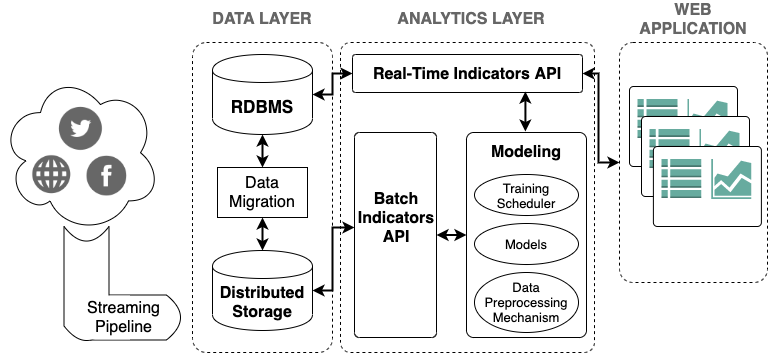}
  \caption{SciLens News Platform architecture. 
  First, a streaming pipeline acts as the entry point of data collection. Then, a data layer, comprised by an RDBMS and a Distributed Storage, stores the incoming data. Lastly, the analytics layer manages the data, trains the Machine Learning models and serves the extracted indicators to the web application.
  }
  \label{fig:arch}
\end{figure}

\subsection{Architecture}
\label{par:architecture}
The architecture of the Scilens News Platform consists of three components which are responsible for the collection, storage and segmentation of data as well as for the models training and the indicators serving to the web application.  
Bellow we describe these components which are also shown in Figure~\ref{fig:arch}.

\textbf{Data Collection and Storage.} 
The SciLens News Platform uses a hybrid data storage scheme that supports both real-time computational operations (with an RDBMS), and ad-hoc querying on historical data and efficient data warehousing (with a Distributed Storage).
The main data entry point of the system is an outlet-based streaming pipeline wrapped around the Datastreamer API (\url{http://www.datastreamer.io}). 
This subsystem acts as a messaging queue and fetches, in real-time, postings from a specific set of social media accounts along with their reactions. 
These incoming data streams are processed, and the corresponding news articles are extracted. 
For these transformations, the system leverages the distributed file system of Hadoop (\url{http://hadoop.apache.org}), and the distributed computational framework Spark (\url{http://spark.apache.org}), for parallel data processing and storing. 
The data synchronization between the RDBMS and the Distributed Storage is made through a daily data migration process. 

\textbf{Data Management and Model Training.}
As we show in our use-case (\S\ref{par:covid}), an essential aspect of our system is the computation of analytics on top of particular segments of our data.
These segments are combinations of content-based supervised topics of news and quality-based categories of news outlets.

More specifically, regarding the content-based segmentation, the system performs a probabilistic hierarchical clustering on the articles and assigns one or more topics to each one of them. 
These topics can be very generic (e.g., Health) or very specific (e.g., COVID-19). 
On the other hand, regarding the outlet quality-based segmentation, the system groups the articles based on the news outlet that they are published and then groups the outlets that have similar quality.
The quality of an outlet is either computed using the expert reviews or imported from external sources (e.g., in \S\ref{par:covid} we use a ranking published by the American Council on Science and Health). 

Finally, our system periodically trains Machine Learning models on top of the Distributed Storage, accessing the full history of our data. 
These models are used for the extraction of the quality indicators that we described in \S\ref{par:indicators}. 


\textbf{Indicators API.} 
The last core component of our system is the Indicators API, which is responsible for the real-time article evaluation. 
Its architecture is based on micro-services, which are lightweight, loosely coupled services that support parallel execution.
The main functionality of this component is to compute and serve quality indicators of articles to the web application.

\section{Demonstration Plan}
\label{par:covid}
Our demonstration will showcase the SciLens News Platform on the topic of the pandemic of \textit{Coronavirus Disease 2019} (\textit{COVID-19}). \textit{COVID-19} is a topic with highly trending nature which triggers an abundance of news articles and social media discussions. 
Given such a prominent topic, the task of discerning between low and high-quality articles becomes very challenging for non-experts in the fields of medicine and epidemiology. 
On that end, we present how fused information retrieved from our system allows end-users to i) assess the quality of individual news articles, and ii) obtain aggregated insights for the topic of \textit{COVID-19}.

To prepare the \textit{COVID-19} data segment, the system uses a shortlist, published by the American Council on Science and Health \cite{alex2017berezow}, that contains 45 mainstream news outlets accompanied by their quality ranking.
The time frame of the data collection, in the context of this paper, covers the 60-day period from 2020-01-15 to 2020-03-15.
A live demonstration of the platform, with up-to-date news articles and quality indicators, can be found here: \url{http://scilens.epfl.ch}.

\begin{figure}[t]
  \centering
  \includegraphics[trim={0 1.5 0 0},clip,width=\linewidth]{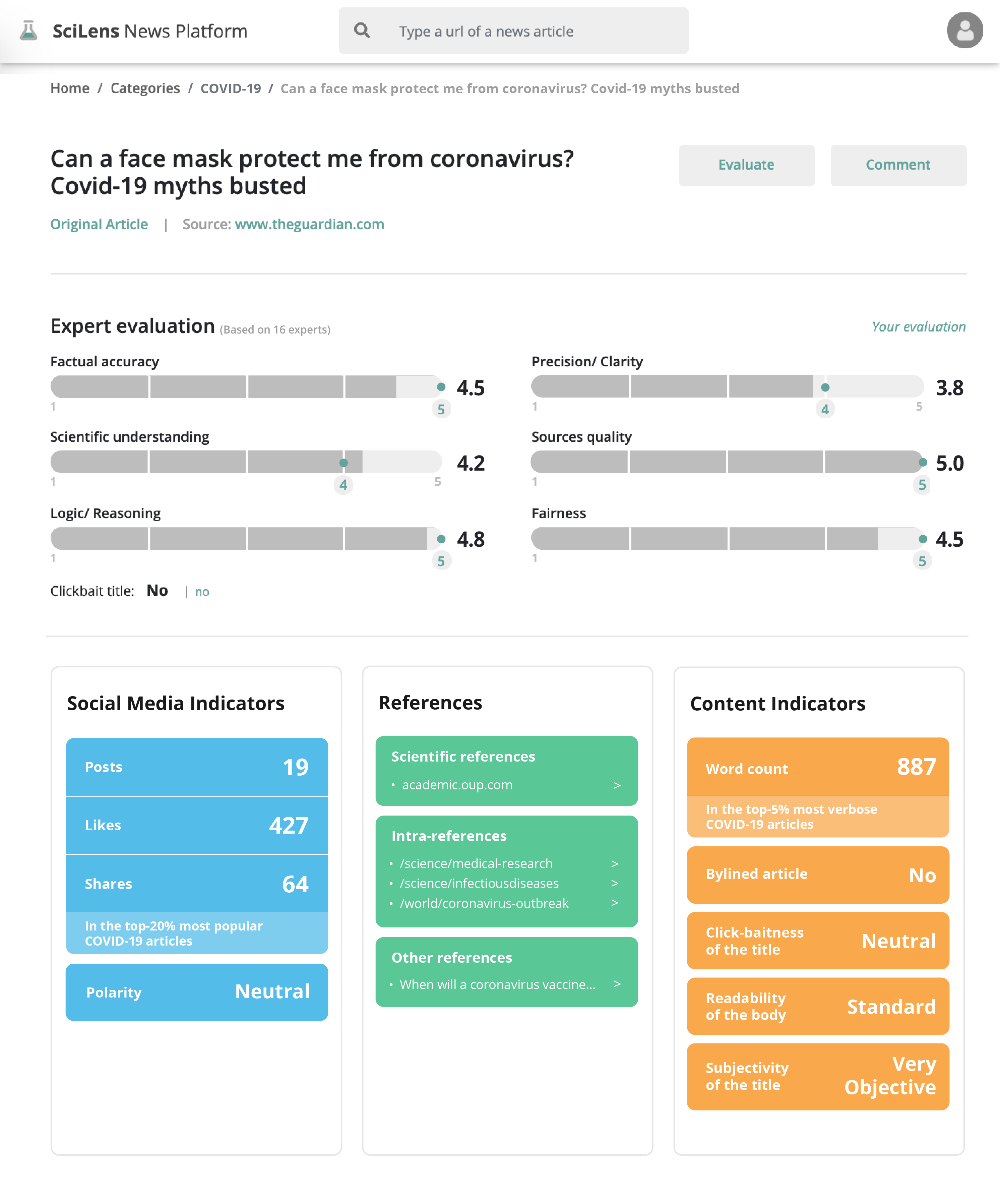}
  \caption{User interface of the SciLens News Platform. A wide range of automatically extracted quality indicators combined with manually-operated expert reviews.}
  \label{fig:ui}
\end{figure}

\subsection{Single Article Assessment}
As we explain in \S\ref{par:overview}, an end-user of the platform can explore in real-time, a wide range of automatically extracted quality indicators combined with manually-operated expert reviews.
A snapshot of this enhanced view of news articles is depicted in Figure~\ref{fig:ui}.
This functionality is available for all the articles in our news collection as well as for any arbitrary news article that a user wants to evaluate.

\subsection{News Topic Insights}
Apart from the single article assessment, a user can interact with aggregated insights regarding a news topic (in our case \textit{COVID-19}).
The news outlets that publish articles regarding \textit{COVID-19} are evaluated based on three axes, namely their newsroom activity, evidence seeking and social engagement.

\begin{figure*}[t]
  \centering
  \includegraphics[width=\linewidth]{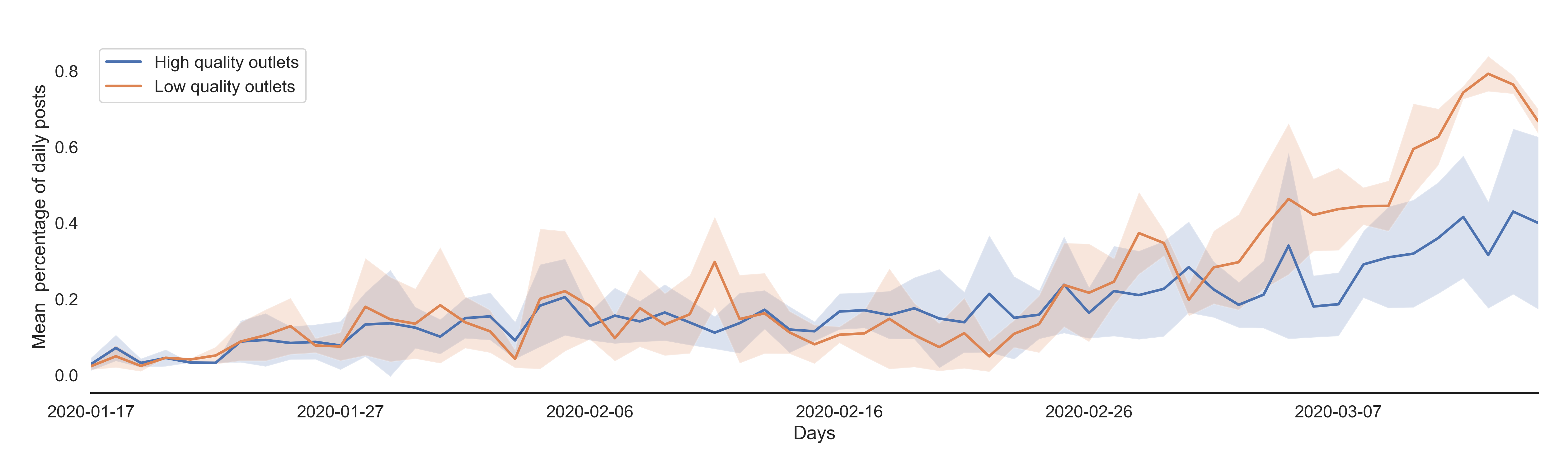}
  \caption{Mean percentage of daily posts referred to COVID-19 per rating category. Low-quality outlets seem to be driven by the breaking news, whereas high-quality outlets are more conservative on their publication rate.}
  \label{fig:daily}
\end{figure*}

\begin{figure}[t]
  \centering
  \includegraphics[width=\linewidth]{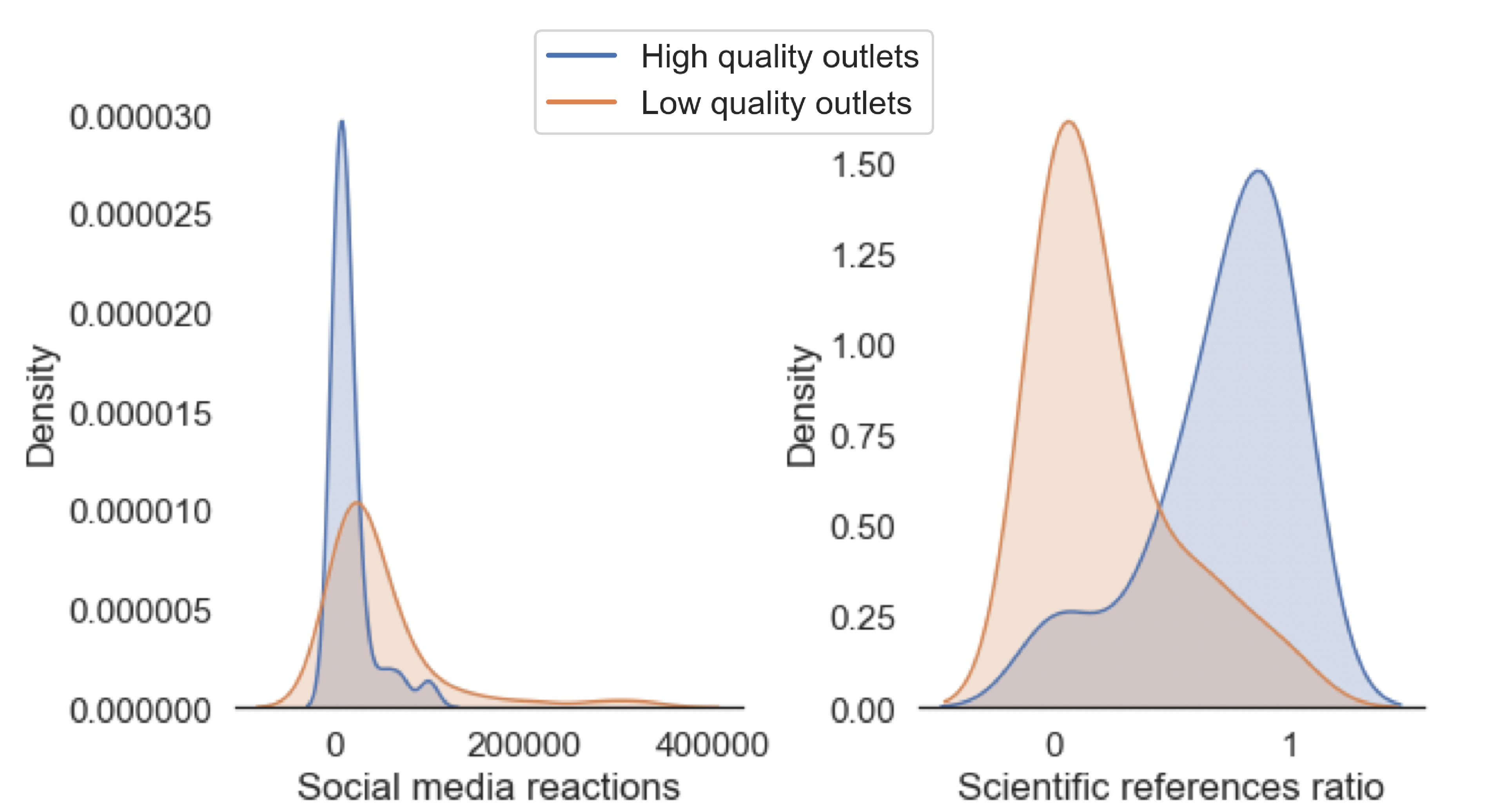}
  \caption{Kernel Density Estimation (KDE) of the number of Social Media Reactions (left) and Scientific References Ratio (right). The low-quality outlets tend to have a wider distribution of reactions but lower number of scientific references, whereas the high-quality outlets show the inverse behaviour.}
  \label{fig:reach}
\end{figure}

To study the newsroom activity, the system computes the distribution of daily posts for each of the outlets.
Then, it groups all the media outlets by their quality ranking and creates a time-series of the mean percentage of daily posts per rating class. The results an end-user will see are presented in Figure~\ref{fig:daily}.

We observe that in the early stages of the discussion on the pandemic, both low and high-quality outlets posted with the same frequency.
However, by the end of the first month, low-quality outlets started dedicating a larger percentage of their published articles on this topic. 
The latter implies a trade-off between the quantity and the quality of the articles. 
Low-quality outlets seem to be driven by the breaking news, whereas high-quality outlets are more conservative on their publication rate; however, as we see next, they have better scientific foundations. 

Moreover, the system provides the end-user with insights regarding i) the social engagement (i.e., the number of social media reactions), and ii) the evidence seeking (i.e., in our use-case, the ratio of scientific references used) of the news outlets.
As shown in Figure~\ref{fig:reach}, one can verify the assumption that low-quality outlets tend to publish more and thus to acquire a higher amount of social media reach than high-quality outlets. 
Conversely, high-quality outlets base their findings more on well-established scientific references.



\balance

\bibliographystyle{abbrvnat}
\bibliography{vldb2020}

\end{document}